\documentclass[leqno,11pt]{article}

\usepackage{amsfonts,latexsym}
\usepackage{amssymb}
\usepackage{amsmath}
\usepackage{amscd}

\usepackage{graphicx}

\DeclareMathOperator{\sech}{sech}
\DeclareMathOperator{\cn}{cn}
\DeclareMathOperator{\sn}{sn}
\DeclareMathOperator{\dn}{dn}

\newcommand{\Sfrac}[2]{{ \textstyle \frac{#1}{#2}}}

\newcommand{\R}{\mathbb{R}}

\DeclareFontFamily{OT1}{pzc}{}
\DeclareFontShape{OT1}{pzc}{m}{it}{<-> s * [1.10] pzcmi7t}{}
\DeclareMathAlphabet{\mathpzc}{OT1}{pzc}{m}{it}
\renewcommand{\a}{\mathpzc{a}}

\numberwithin{equation}{section}

\begin{document}

\title{\Large{\bf Convective Wave Breaking in the KdV Equation}}
\author{Mats K. Brun\footnote{\texttt{Mats.Brun@math.uib.no}}
\ and 
Henrik Kalisch\footnote{\texttt{henrik.kalisch@math.uib.no}} \\
        Department of Mathematics, University of Bergen \\
        P.O. Box 7800, 5020 Bergen, Norway}

\maketitle

\begin{abstract}
The KdV equation is a model equation for waves at the surface of an inviscid
incompressible fluid,
and it is well known that the equation describes the evolution
of unidirectional waves of small amplitude and long wavelength fairly accurately
if the waves fall into the Boussinesq regime.

The KdV equation allows a balance of nonlinear steepening effects
and dispersive spreading which leads to the formation of steady wave profiles
in the form of solitary waves and cnoidal waves.
While these wave profiles are solutions of the KdV equation for any amplitude,
it is shown here that there for both the solitary and the cnoidal waves,
there are critical amplitudes for which the horizontal
component of the particle velocity matches the phase velocity of the wave.
Solitary or cnoidal solutions of the KdV equation which surpass these 
amplitudes feature incipient wave breaking as the particle velocity exceeds 
the phase velocity near the crest of the wave,
and the model breaks down due to violation of the kinematic
surface boundary condition.

The condition for breaking can be conveniently formulated as
a convective breaking criterion based on the local Froude number
at the wave crest. This breaking criterion can also be applied
to time-dependent situations, and one case of interest is
the development of an undular bore created by an influx at
a lateral boundary. 
It is shown that this boundary forcing leads to wave
breaking in the leading wave behind the bore
if a certain threshold is surpassed.
\end{abstract}

\section{Introduction}
The Korteweg-de Vries (KdV) equation is a model equation describing
the evolution of long-crested waves at the surface of a body of fluid. 
The equation is derived as a physical model equation 
under the assumption that there is an approximate
balance between nonlinear steepening effects and 
dispersive spreading. In mathematical terms,
this balance is expressed by introducing two small parameters
$\alpha$ and $\beta$ measuring the wave amplitude and the wave length, respectively.

The main assumptions on the waves to be represented by solutions 
of the KdV equation are that they be of small amplitude
and long wavelength when compared to the undisturbed depth of the fluid.
Suppose that the undisturbed depth of the fluid is given by $h_0$.
If $\a$ represents a typical wave amplitude, and $\ell$ represents a typical
wavelength, the two parameters $\alpha = \a / h_0$ and 
$\beta = h_0^2 / \ell^2$ should be small and of the same order.
This is the Boussinesq scaling. If in addition the wave motion 
is predominantly in a single direction, then the KdV equation
is an approximate model describing the surface wave motion \cite{BCL,LannesBOOK}.
If the undisturbed depth $h_0$ is taken as a unit of distance 
and $\sqrt{h_0/g}$ is taken as a unit of time, then the KdV equation
appears in the form
\begin{equation}
\label{kdv}
\eta_t + \eta_x + \Sfrac{3}{2} \eta \eta_x + \Sfrac{1}{6} \eta_{xxx} = 0.
\end{equation}

In the present article, the focus is on whether -- or rather how -- the KdV
equation is able to describe incipient wave breaking.
While the KdV equation does not admit the distinctive steepening and
development of infinite gradients known from nonlinear hyperbolic equations,
it will be shown here that the KdV equation features a different kind of wave
breaking which is more closely related to spilling at the wave crest.

Let us recall that the KdV equation can be thought of as a combination
of the simple nonlinear balance law
\begin{equation}\label{Burgers}
\eta_t + \eta_x + \Sfrac{3}{2} \eta \eta_x = 0,
\end{equation}
and the linear dispersive equation
\begin{equation}\label{Airy}
\eta_t + \eta_x + \Sfrac{1}{6}\eta_{xxx} = 0.
\end{equation}
While all waves featuring negative slope break at some point
in the model \eqref{Burgers}, no waves break in the model \eqref{Airy}
(however, \eqref{Airy} features dispersive blow-up 
for some data \cite{BonaSaut, KharifPeli, Rauch}).
The KdV equation \eqref{kdv} allows a balance of nonlinear steepening effects
and dispersive spreading which arrests the typical hyperbolic wave breaking
exhibited by \eqref{Burgers}, and leads to the formation of steady
traveling waves, such as solitary and cnoidal waves \cite{BOOK,KdV}
which propagate without a change in the wave profile.

Specifically, solitary-wave solutions of \eqref{kdv} are of the form
\begin{equation} \label{kdvsol}
\eta(x,t) =H \mbox{sech}^2\big( \Sfrac{\sqrt{3H}}{2}(x-x_0-ct)\big),
\end{equation}
where the phase velocity is given by
$c=1+\frac{H}{2}.$
While these formulas define solutions of the KdV equation for all waveheights $H$,
it turns out that solutions of large waveheight are inconsistent with the 
model in which the KdV equation is valid. In particular, as will be shown
in Section 2, for waveheights exceeding the critical waveheight
$H_{\text{max solitary}} = 0.6879$, the phase velocity of the wave
is smaller than the particle velocity at the crest, 
a fact which may be interpreted as incipient wave breaking.

\begin{figure}
  \begin{center}
     {\includegraphics[scale=0.38]{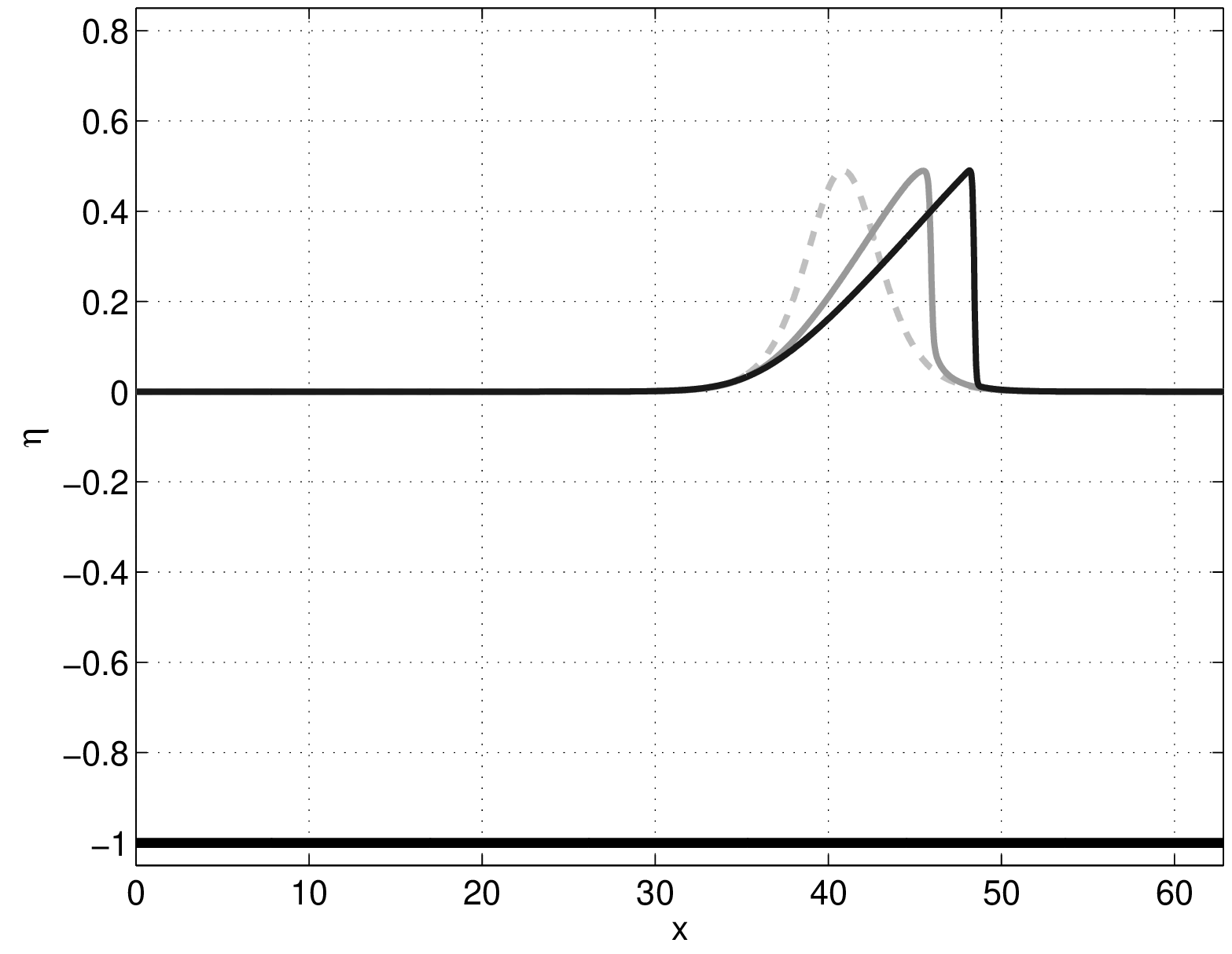}}~~~~
     {\includegraphics[scale=0.38]{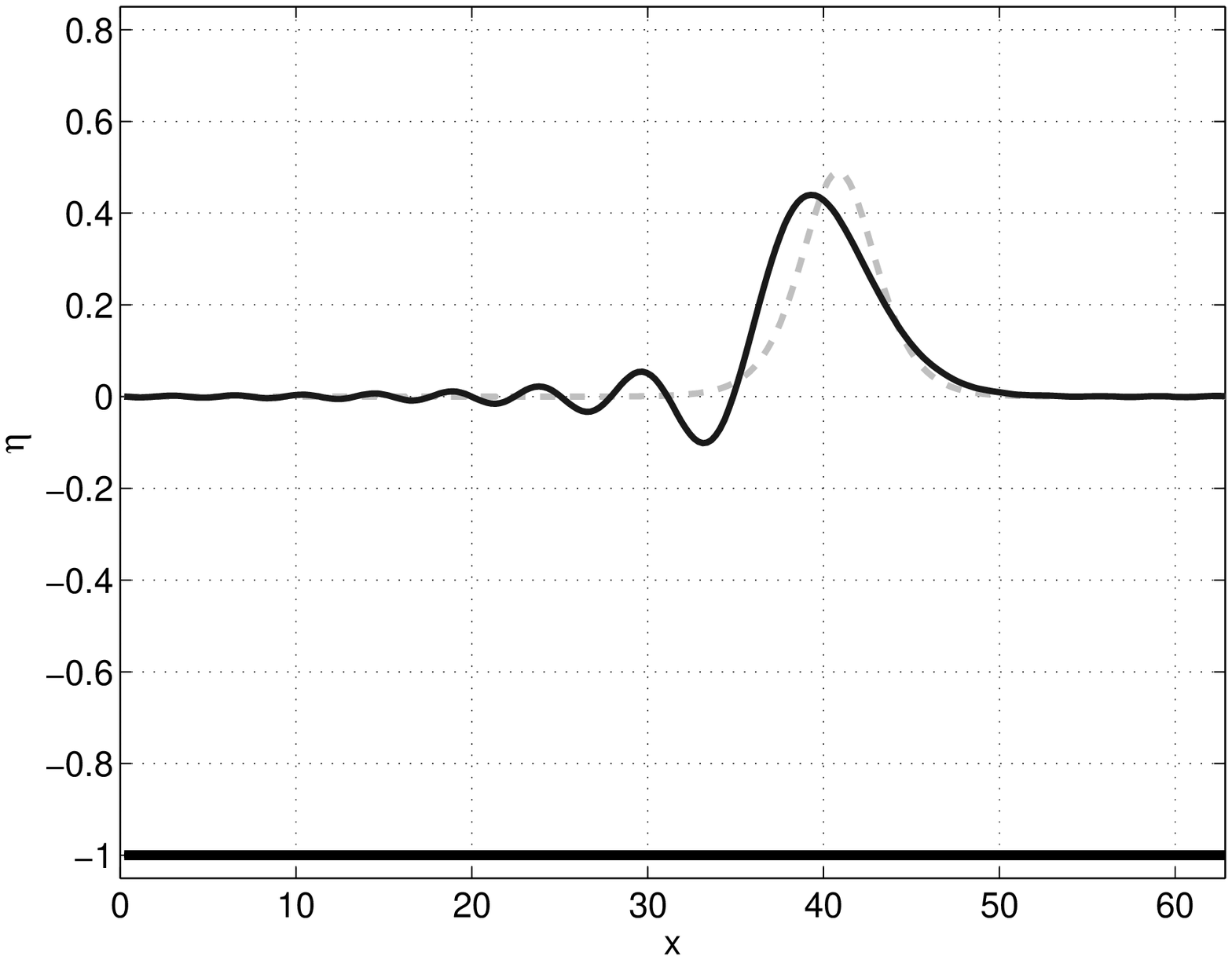}}
  \end{center}
  \caption{\small Left panel: Solution of Burgers equation $\eta_t + \Sfrac{3}{2}\eta \eta_x = 0$ which features
hyperbolic wave breaking. Right panel: Solution of the Airy equation $\eta_t + \Sfrac{1}{6}\eta_{xxx}$
which features dispersive spreading. Initial data are shown as dashed curves. 
If the dashed curve were used as initial data for the KdV equation \eqref{kdv}, 
then this wave would propagate without change in shape.}
\label{NonlinearityDispersion}
\end{figure}

To elaborate on this argument, note that
it is shown in \cite{Wh}, that the derivation of the equation 
as a surface water-wave model enables the reconstruction of
an approximation of the fluid velocity field underneath the surface.
In particular, it is possible to derive relations expressing the horizontal 
and vertical velocity components in terms of the principal unknown variable 
$\eta$ which describes the shape of the free surface.
The horizontal
velocity component in the fluid is given by
\begin{equation} \label{SecOrdRel}
u(x,y,t) = \eta - \Sfrac{1}{4} \eta^2 + \big(\Sfrac{1}{3} - \Sfrac{y^2}{2} \big) \eta_{xx},
\end{equation}
and this relation holds to the same order in the asymptotic parameters
$\alpha$ and $\beta$ as to which the KdV equation is valid.

Since the horizontal component of the particle velocity can be 
found approximately, it may be compared to the local phase velocity
of the wave. This leads to one of the most fundamental breaking criteria 
used in the literature, the convective breaking criterion 
which predicts wave breaking if the particle velocity at the wavecrest 
exceeds the phase velocity of the wave (see for example \cite{LH1969,TL1992}).
More specifically, this breaking criterion (often termed kinematic criterion) 
is based
on the local Froude number, and predicts breaking if
\begin{equation}\label{break1}
\frac{U}{C} \geq 1,
\end{equation}
where $U = u(x,\eta(x,t),t)$ is the horizontal component of the velocity field evaluated
at the free surface, $C$ is the local phase velocity of the wave.
To illustrate this point of view, Figure 2 shows two waves with 
corresponding fluid particles near the crest of the wave.
In the left panel, a wave with sufficiently small amplitude is depicted. A fluid particle located near the
wavecrest remains in the free surface and recedes from the crest in unison with the wave motion. 
However in the right panel, a wave of higher amplitude
is shown, and particles located near the crest have higher horizontal velocity than the wave itself.

The utility of the convective criterion has been the subject
of some discussions in the literature. 
For wave breaking in shallow water, such as in focus in the present article,
some studies such as \cite{Nepf}
conclude that the convective criterion
does a fair job of predicting wave breaking even for three-dimensional waves.
On the other hand, studies focusing on breaking of deep-water waves
such as \cite{Perlin} observe
that the kinematic criterion has been useful in some practical situations,
while there is also evidence that it may not be a reliable indicator 
to pinpoint the onset of breaking in the case of deep water.
In some of the cases where the convective breaking criterion performs badly, 
the difficulty of obtaining accurate readings for the 
phase velocity of the wave in experimental situations may be responsible \cite{Stansell}.

Note that the criterion \eqref{break1} is stated in sufficient generality to be applicable
to both experimental and numerical work.
In the case of numerical modeling, and in particular in the case of phase-resolving models,
there is no difficulty in evaluating the local wave and particle velocities.
The local phase velocity can usually be be approximated using a variety of methods.
For example, Fourier techniques have been used in \cite{Sato}.
What is more, in the case of solitary and cnoidal waves, the phase and particle velocities
are known in closed form, and it is straightforward to test the breaking criterion \eqref{break1}.
These computations will be carried out in sections 2 and 3.

While it is relatively easy to evaluate the breaking criterion in
several situations for the equation \eqref{kdv},
it should also be noted that
free surface waves in shallow water are not likely to feature breaking unless
some focusing effect is present, such as an underlying current 
\cite{Perlin,YaoWu}
or strong three-dimensional effects \cite{Nepf}. 
On the other hand wave breaking is likely
to occur in the presence of forcing in the form of an uneven bottom topography
or a discharge. These two possibilities are sketched in Figure \ref{NaturalBreaking}.
In particular the left panel of Figure \ref{NaturalBreaking} indicates
one of the most widely known type of wave breaking, the development of
breakers on a sloping beach. In this case, depending on a number of parameters such as waveheight, wavelength and bottom slope,
a variety of breaking phenomena takes place,
ranging from spilling at the crest, to overturning and surging breaking \cite{DD}.

In the right panel of Figure \ref{NaturalBreaking},
the case of a forced inflow is indicated.
In this case, a so-called bore generally emerges. 
Sufficiently small discharges lead to the appearance of an undular bore
which is a characterized by a moderately steep front followed by a 
leading wave and trailing smaller waves.
For larger inflows, the leading wave behind the bore may break, and
for large enough inflows, the bore can be entirely turbulent \cite{Chanson,KochCha}.
In the final section of the present paper, the case of a forced inflow 
and undular bore is discussed in some detail, and the question of the
transition from purely undular to undular with spilling breaking
is investigated.

\begin{figure}
  \begin{center}
     {\includegraphics[scale=0.38]{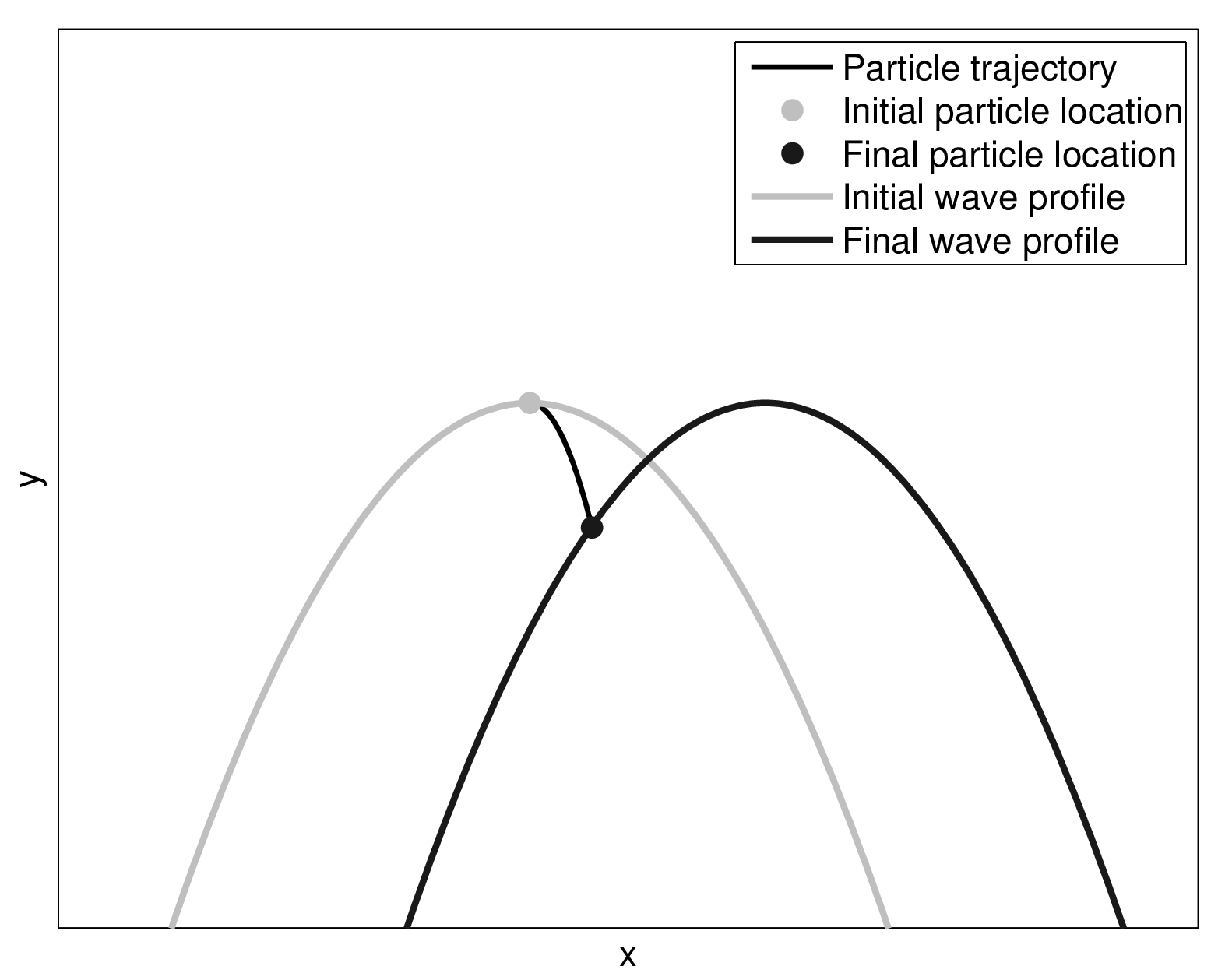}}~~~~
     {\includegraphics[scale=0.38]{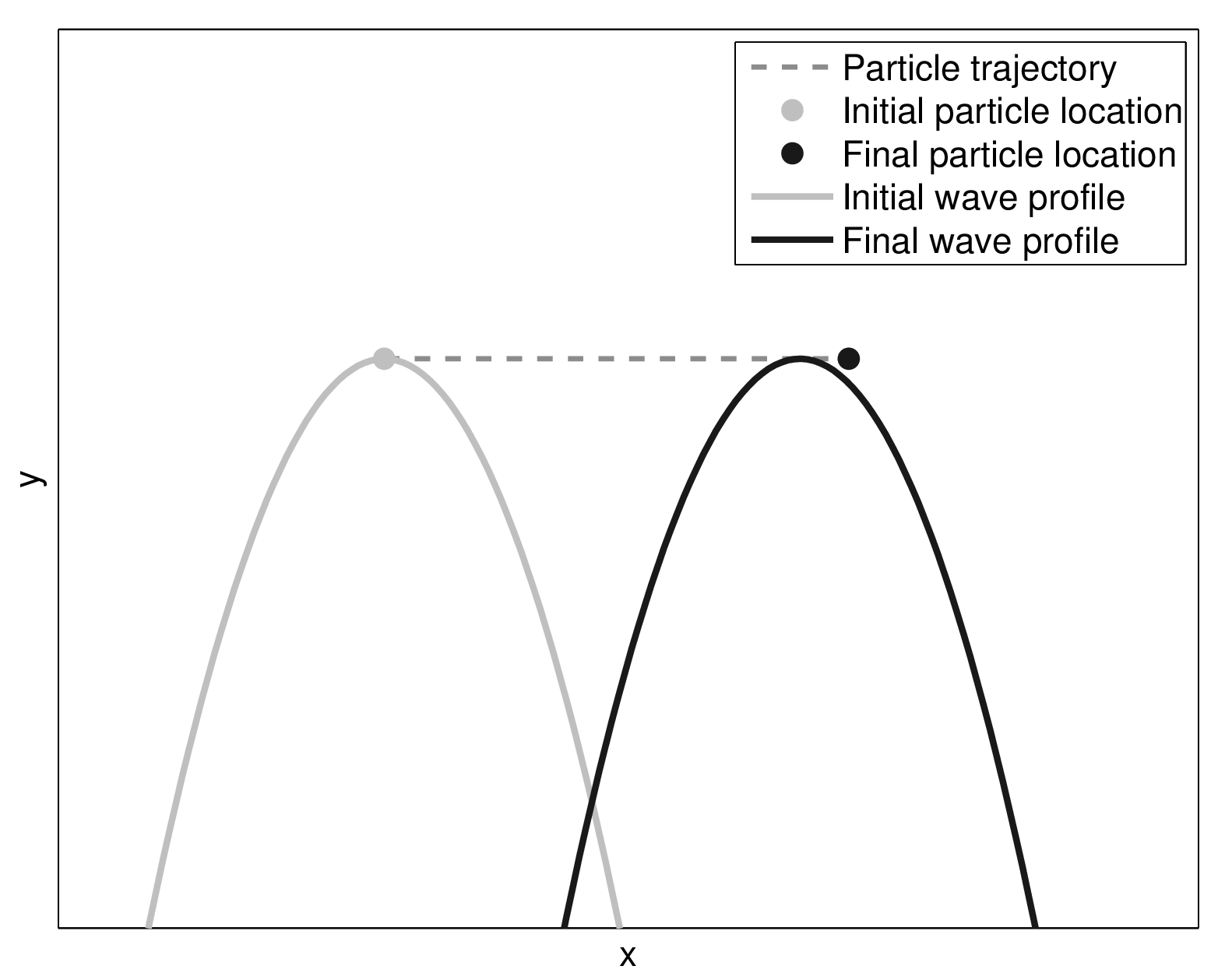}}
  \end{center}
  \caption{\small Left panel: Surface profile of a regular wave, and trajectory of
a particle in the free surface.
Right panel: Surface profile of a breaking wave. A particle contained in the free surface
is leaving the fluid domain, indicating incipient wave breaking.}
\label{Bore_Picture}
\end{figure}

Before we enter the main development of this paper, some remarks are in order.
First, it should be noted that
convective wave breaking of the type discussed here 
is unlikely to happen in hyperbolic equations or systems.
While the simple conservation law $\eta_t + \frac{3}{2}\eta \eta_x = 0$
does not contain expressions for the local particle velocity, the
nonlinear shallow-water equations
\begin{align*}
h_t + (uh)_x = 0, \\
u_t + h_x + uu_x = 0,
\end{align*}
include both the surface deflection $\eta(x,t) = h(x,t)-1$, and an average horizontal particle
velocity $u(x,t)$ in the description.
In this case, a simple wave propagating to the right is given by the
Riemann invariant $u+2 \sqrt{h}$, and has phase velocity $u+\sqrt{h}$.
As shown in \cite{Wh}, for a wave profile $h=H(x)$ with undisturbed depth $1$, 
the fluid velocity is $u = 2 \sqrt{H}-2$, and the phase velocity of the wave is
$C = 3 \sqrt{H}-2$. 
Thus we see that convective breaking does not happen in this case.
On the other hand, the typical hyperbolic steepening and eventual breaking
will happen for appropriate initial conditions.

\begin{figure}
  \begin{center}
  {\includegraphics[scale=0.4]{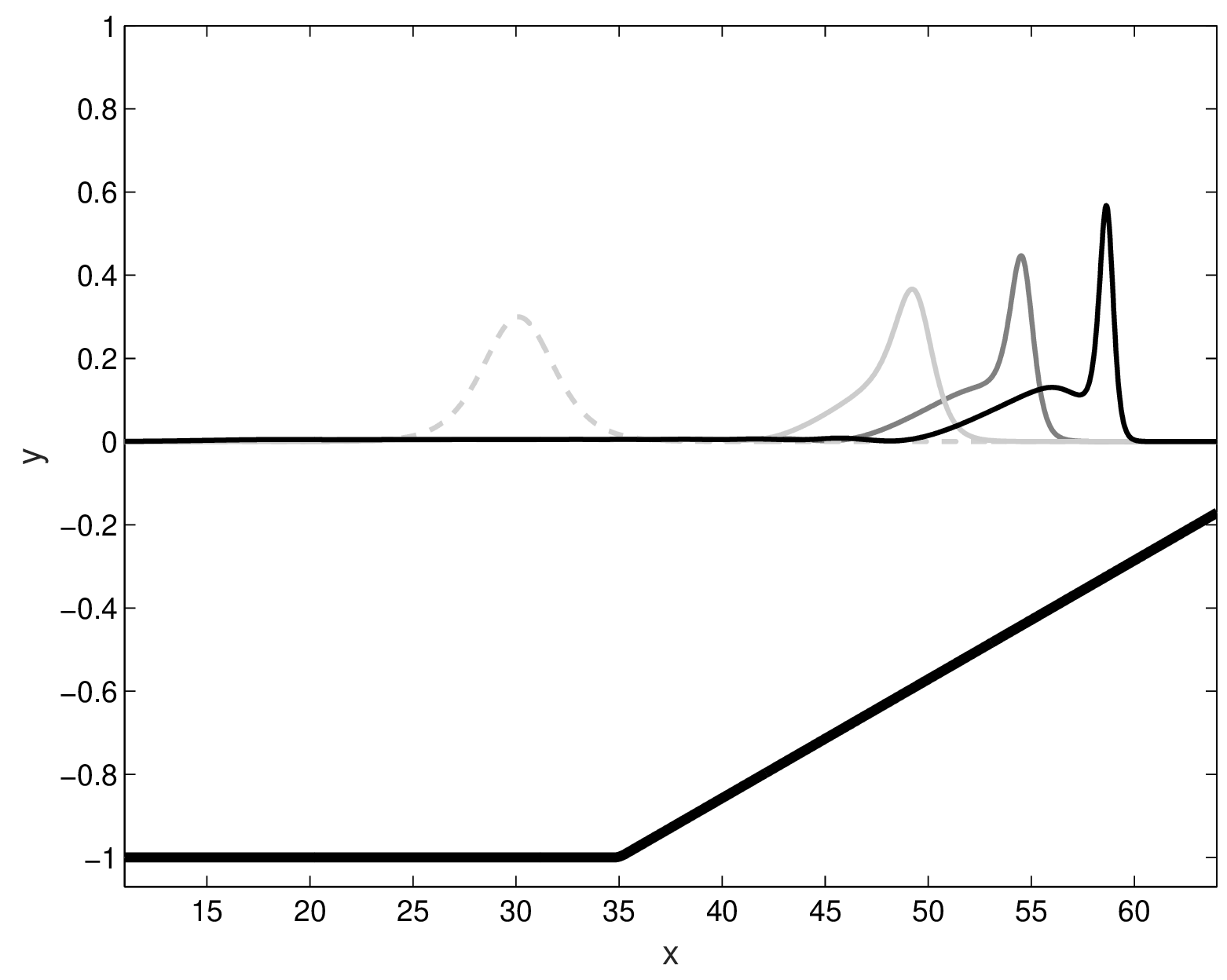}}~~~~~~
  {\includegraphics[scale=0.4]{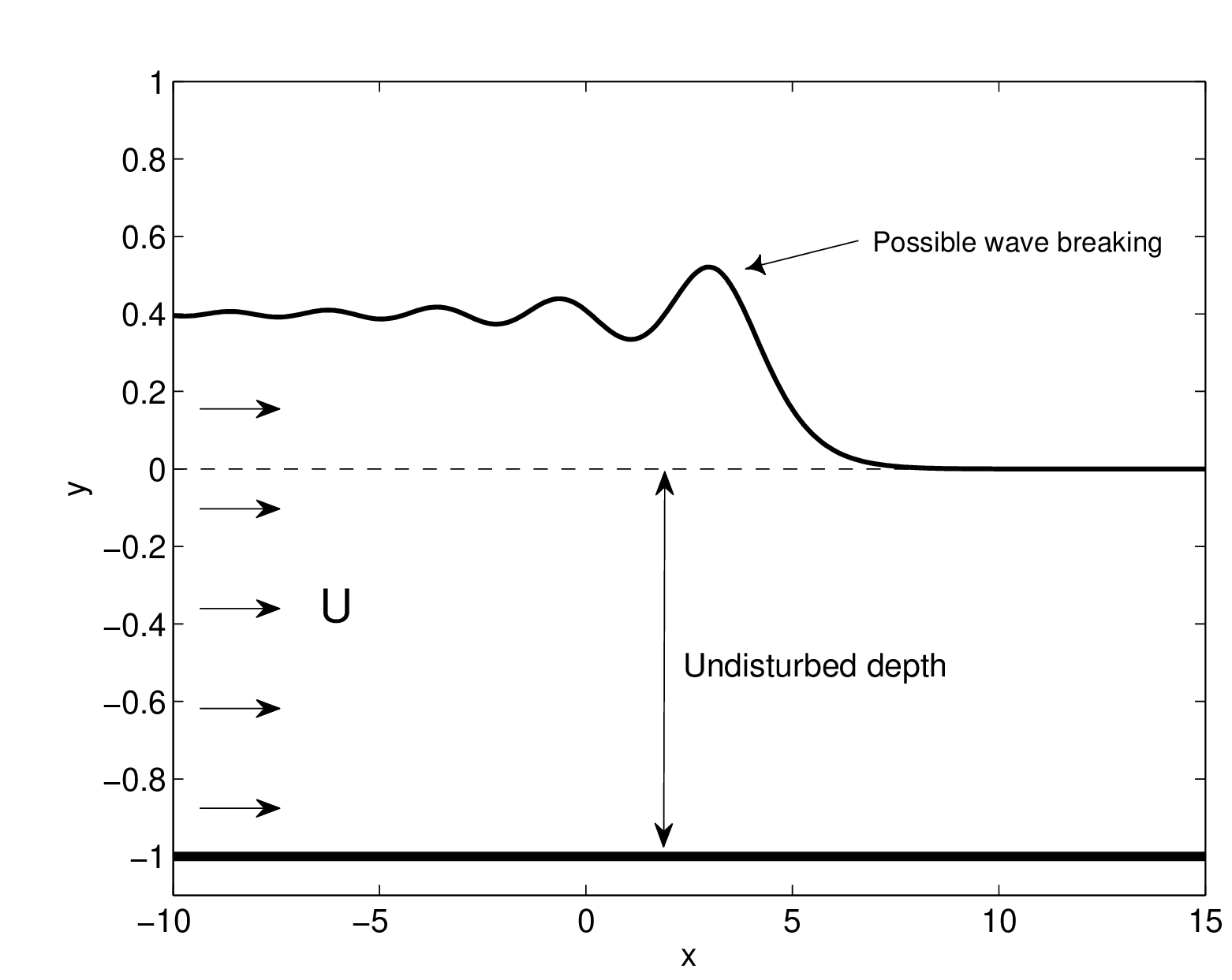}}
  \end{center}
  \caption{\small Left panel: Breaking wave in Boussinesq model with sloping bottom.
Right panel: Wave breaking due to a large discharge at left boundary.}
\label{NaturalBreaking}
\end{figure}

Secondly, it should be mentioned that
the relation \eqref{SecOrdRel} and similar expressions for the vertical velocity
may also be used advantageously for a number of purposes, such as 
the definition of a family of Boussinesq models \cite{BCS1},
describing particle paths beneath the surface \cite{BoKa},
and and the study of 
mechanical balance laws associated to the evolution equation \cite{AK1,AK3}.
On the other hand,
there is a variety of different strategies to construct the velocity field in the fluid,
such as the analytic method used in \cite{He} to find the velocities
associated to a periodic cnoidal solution of the KdV equation.
Particle paths under periodic traveling waves can also be found
to rather high accuracy using the Lagrangian approach \cite{ChHsCh}.
Concerning the notion of mechanical balance laws, we mention in particular that 
an analysis along these lines casts doubt
on the conservation of mechanical energy in the KdV approximation \cite{AK3}.
The conservation of energy in both the KdV and higher-order KdV equations
has also been questioned recently in a study which utilizes a Lagrangian framework \cite{KRI2015}.

The plan of the paper is as follows. In Section 2, the convective breaking criterion is applied
to the solitary wave, and it is found that the limiting waveheight is $H_{\text{max solitary}} = 0.6879$.
In Section 3, the breaking criterion is applied to the cnoidal wave solutions. It is found
that cnoidal wave solutions may feature breaking at a much smaller waveheight than the solitary wave.
Finally, Section 4 is devoted to the study of wave breaking in a dynamic situation with a forced
inflow. The resulting undular bore features a leading wave which may break, and
the focus is on finding the critical case dividing the field of purely undular bores from
partially turbulent breaking bores.

\section{Convective wave breaking for solitary waves}
\label{sec:breakingcrit}

The surface water-wave problem for an inviscid fluid
is generally described by the Euler equations with an impenetrable bottom, 
and  kinematic and dynamic boundary conditions at the free surface. 
Assuming weak transverse effects, the unknowns
are the surface elevation $\eta(x,t)$, the horizontal and vertical 
fluid velocities $u(x,y,t)$ and $w(x,y,t)$, respectively, and the pressure $P(x,y,t)$.
If the assumption of irrotational flow is made, then
a velocity potential $\phi(x,y,t)$ can be used.
In the non-dimensional variables mentioned in the introduction, 
the problem may be posed on the domain
$\left\{(x,y) \in \R^2 \, | \, 0 < y < 1+\eta(x,t) \right\}$.
The derivation of the KdV equation is based on approximating the velocity potential
by an asymptotic series. In suitably scaled variables, 
the horizontal velocity component takes the form
\begin{equation}\label{w}
u = \phi_x = w - \beta \Sfrac{y^2}{2} w_{xx} + O(\beta^2),
\end{equation}
where $w(x,t) $ represents the horizontal velocity at the bottom.
On the other hand, the one-way assumption inherent in the KdV equation
necessitates that the velocity at the bed is given by
\begin{equation*}
w = \eta - \Sfrac{1}{4} \alpha \eta^2 + \Sfrac{1}{3} \beta \eta_{xx} + O(\alpha^2, \alpha\beta, \beta^2).
\end{equation*}
By combining these two relations and neglecting terms of $O(\alpha^2, \alpha \beta, \beta^2)$, 
we obtain an expression for the horizontal component of the velocity field in terms of $\eta$ and $y$:
\begin{equation*}
u = \phi_x = \eta - \Sfrac{1}{4} \alpha \eta^2 + \beta \big(\Sfrac{1}{3} - \Sfrac{y^2}{2}\big)\eta_{xx}.
\end{equation*}
In the original variables, 
the horizontal velocity is then expressed as
\begin{equation}\label{uab}
u(x,y,t) = \eta - \Sfrac{1}{4} \eta^2 + \big(\Sfrac{1}{3} - \Sfrac{y^2}{2}\big) \eta_{xx},
\end{equation} \\
and the breaking criterion \eqref{break1} can be written as
\begin{equation}\label{break2}
\eta - \Sfrac{1}{4} \eta^2 + \big(\Sfrac{1}{3} - \Sfrac{(1 + \eta)^2}{2}\big) \eta_{xx} \geq C.
\end{equation}

We now apply this breaking criterion to the solitary wave given by \eqref{kdvsol},
where $H$ is the waveheight, $c$ is the wavespeed, 
and $x_0$ is the initial location of the wave crest.
Differentiating twice with respect to $x$ yields
\begin{equation*}
\eta_{xx} = 3 H^2 \sech^2 \big(\Sfrac{\sqrt{3 H}}{2} (x - ct - x_0) \big) 
                \tanh^2 \big( \Sfrac{\sqrt{3 H}}{2} (x - ct - x_0) \big) 
        - \Sfrac{3}{2} H^2 \sech^4 \big(\Sfrac{\sqrt{3 H}}{2} (x - ct - x_0) \big).
\end{equation*}
Since the solitary wave retains its shape for all time we may evaluate at 
$x-ct-x_0 = 0$, in which case we obtain
$\eta = H$ and $\eta_{xx} = -\Sfrac{3}{2} H^2$.
Substituting these values into equation \eqref{uab} and evaluating at the surface
yields the breaking criterion \eqref{break1}
\begin{equation}
H - \Sfrac{1}{4} H^2 - \Sfrac{3}{2} \big( \Sfrac{1}{3} - \Sfrac{(1 + H)^2}{2} \big) H^2 \geq 1 + \Sfrac{1}{2}H .
\end{equation}
Setting the left hand side equal to the right hand side and rearranging terms yields
\begin{equation}\label{solpol}
\mathcal{P}(H) = \Sfrac{3}{4} H^4 + \Sfrac{3}{2} H^3 + \Sfrac{1}{2} H - 1 =  0
\end{equation}
for the critical value of the waveheight.
We see that $\mathcal{P}$ is a fourth order polynomial in $H$, 
and since $ \mathcal{P}'(H) = 3 H^3 + \frac{9}{2} H^2 + \frac{1}{2} > 0$ for $H \geq 0$  
and $\mathcal{P}(0) < 0$ while $ \mathcal{P}(1) > 0$, 
it can have only one positive root which lies in $[0,1]$.
This root can be found numerically 
to obtain the following value of the maximum allowable waveheight for the solitary wave:
\begin{equation}
H_{\text{max solitary}} = 0.6879.
\end{equation}
Even though the KdV equation admits solutions with any waveheight, waves with a
waveheight larger than $H_{\text{max solitary}}$ do not describe actual surface waves
since these waves already feature incipient wave breaking.
Of course the waveshape of a solitary-wave solution of the KdV equation
may not be close to a surface water wave already for smaller waveheights 
than $H_{\text{max solitary}}$ (cf. \cite{BKN}).

\section{Maximum waveheight for periodic waves}
The cnoidal waves are defined with the help of the
incomplete elliptic integral of the first kind.
For a given elliptic parameter $0<m<1$, this integral is
\begin{equation*}
u = \int_0^{\phi} \Sfrac{d\theta}{\sqrt{1 - m\sin^2\theta}}.
\end{equation*}
The elliptic functions 
$\sn(\cdot | m)$, $\cn(\cdot | m)$ and $\dn(\cdot |m)$ are defined as 
\begin{equation*}
\begin{split}
\sn(u | m) &= \sin \phi, \\
\cn(u | m) &= \cos \phi, \\
\dn(u |m) &= \sqrt{1 - m\sin^2\phi}.
\end{split}
\end{equation*}
On the other hand, the complete elliptic integrals of the first and second kind, respectively, are defined by
\begin{equation*}
K(m) = \int_0^{\pi/2} \Sfrac{d\theta}{\sqrt{1 - m\sin^2\theta}} \  \ \ \text{ and } \ \ E(m) = \int_0^{\pi/2}\sqrt{1 - m\sin^2\theta} \ d\theta.
\end{equation*} \\
As shown in \cite{Dingemans},
the cnoidal travelling wave solutions of \eqref{kdv} are given 
in terms of three parameters $f_1$, $f_2$ and $f_3$ by
\begin{equation}\label{cnsol}
\begin{split}
\eta(x,t) &= f_2 + (f_1 - f_2) \cn^2 \Big( \Sfrac{\sqrt{3(f_1 - f_3)}}{2} (x - ct - x_0) \big| m \Big),
\end{split}
\end{equation}
where $H = f_1 - f_2$ is the wave height, $c = 1 + \frac{1}{2}(f_1 + f_2 + f_3)$
is the wavespeed, and $m=(f_1-f_2)/(f_1-f_3)$ is the elliptic parameter. 
Any cnoidal wave is completely determined as long as these three constants are fixed.
If $\sigma$ is defined by $\sigma^2 = \frac{4}{3(f_1 - f_3)}$,
then the wavelength can be written as $\lambda = 2 \sigma K(m)$.

Since the wave oscillates about the equilibrium level of the surface, 
the mean value of the surface displacement over one wavelength should be zero. 
This requirement can be expressed by integrating over one wavelength, viz. 
$$
\int_0^{\lambda} \eta \, dx = 0,
$$
and it can be shown that this integral is
\begin{equation}
\int_0^{\lambda} \eta \, dx =
2 \sigma \Big( (f_1 - f_3) E(m) + f_3 K(m) \Big)
  = 2 \sigma \Big( \Sfrac{f_1-f_2}{m}E(m) + f_3 K(m) \Big).
\end{equation}
Since $H = f_1 - f_2$, we therefore have  $f_3 = -\frac{H}{m} \frac{E(m)}{K(m)}$.
The parameters $f_1$ and $f_2$, can then also be expressed in terms of wave height $H$ 
and the elliptic parameter $m$ as follows:
\begin{equation*}
\begin{split}
 f_1 &= \Sfrac{H}{m} \left(1 - \Sfrac{E(m)}{K(m)}\right), \\ 
 f_2 &= \Sfrac{H}{m}\left(1 - m - \Sfrac{E(m)}{K(m)}\right).
\end{split}
\end{equation*}
Thus, fixing $H$ and $m$ is sufficient to specify any cnoidal wave, 
as long as the undisturbed water level is set to zero.

To determine whether a given cnoidal wave is a reasonable description of a water wave
in the KdV approximation, we test the wave breaking criterion \eqref{break2}.
By differentiating \eqref{cnsol} twice with respect to $x$ we obtain
\begin{equation*}
\begin{split}
\eta_{xx}  = & \Sfrac{3}{2} (f_1 - f_2)(f_1 - f_3) \Big\{ \sn^2 \big(\Sfrac{\zeta}{\sigma}|m \big) \dn^2 \big( \Sfrac{\zeta}{\sigma} |m \big) + \\
&m^2 \sn^2\left(\Sfrac{\zeta}{\sigma} |m\right) \cn^2\left(\Sfrac{\zeta}{\sigma} |m\right) - \cn^2\big(\Sfrac{\zeta}{\sigma} |m\big) \dn^2\big(\Sfrac{\zeta}{\sigma}|m \big) \Big\},
\end{split}
\end{equation*}
where the argument is given by $\zeta = x-ct-x_0$.
Evaluating at $\zeta = 0$ yields
\begin{equation*}
\begin{split}
\eta &= f_1, \\
\eta_{xx} &= -\Sfrac{3}{2}(f_1 - f_2)(f_1 - f_3).
\end{split}
\end{equation*}
Substituting these expressions into equation \eqref{break2} transforms the breaking criterion 
\eqref{break2} into
\begin{equation}\label{cnbreak}
f_1 - \Sfrac{1}{4}f_1^2 - \Sfrac{3}{2}\left(\Sfrac{1}{3} - \Sfrac{(1 + f_1)^2}{2}\right)(f_1 - f_2)(f_1 - f_3) \geq 1 + \Sfrac{1}{2}(f_1 + f_2 + f_3).
\end{equation}
Defining the constants
\begin{equation*}
a = \Sfrac{1}{m}\left(1 - \Sfrac{E(m)}{K(m)}\right), \ b = \Sfrac{1}{m} \left(1 - m - \Sfrac{E(m)}{K(m)}\right) \ \text{ and } \ c = \Sfrac{1}{m} \Sfrac{E(m)}{K(m)},
\end{equation*}
we can write the constants  $f_1$, $f_2$ and $f_3$ as
\begin{equation*}
f_1 = Ha, \ f_2 = Hb \ \text{ and } \ f_3 = -Hc.
\end{equation*}
Substituting these definitions into equation \eqref{cnbreak} 
and setting the left and right hand side equal yields
\begin{equation}\label{cnpol}
\begin{split}
Q_m(H) = \frac{3}{4} &H^4(a^4 + a^3c - a^3b - a^2bc) \\
+ \ \frac{3}{2}&H^3(a^2c + a^3 - a^2b - abc) \\
+ \ \frac{1}{4}&H^2(ac - bc - ab) \\
+ \ \frac{1}{2}&H(a - b + c) - 1 = 0.
\end{split}
\end{equation}
%
%
%
\begin{table}
\begin{center}
  \begin{tabular}[]{l c c c c c}
\hline\hline
$m$ & $H_{\text{max cnoidal}}$ & $\lambda$ & $\alpha$ & $\beta$ & $\mathcal{S}$\\
\hline
0.01 & 0.0196 & 2.591 & 0.0098 & 0.1489 & 0.0661 \\
\vspace{0.05cm}
0.1 & 0.1698 & 2.857 & 0.0849 & 0.1224 & 0.6933 \\
\vspace{0.05cm}
0.2 & 0.2909 & 3.178 & 0.1454 & 0.0990 & 1.4689\\
\vspace{0.05cm}
0.3 & 0.3820 & 3.507 & 0.1910 & 0.0812 & 2.3499\\
\vspace{0.05cm}
0.4 & 0.4548 & 3.849 & 0.2274 & 0.0674 & 3.3702\\
\vspace{0.05cm}
0.5 & 0.5152 & 4.218 & 0.2575 & 0.0562 & 4.5834\\
\vspace{0.05cm}
0.6 & 0.5667 & 4.632 & 0.2833 & 0.0465 & 6.0813 \\
\vspace{0.05cm}
0.7 & 0.6114 & 5.128 & 0.3056 & 0.0380 & 8.0399\\
\vspace{0.05cm}
0.8 & 0.6504 & 5.781 & 0.3252 & 0.0299 & 10.869\\
\vspace{0.05cm}
0.9 & 0.6841 & 6.829 & 0.3420 & 0.0214 & 15.952\\
\hline \hline
\end{tabular}
\caption{\small Critical waveheight for the cnoidal solution of the KdV-equation, calculated for various values of the elliptic parameter $m$. Corresponding values of wavelength and the dimensionless parameters $\alpha$ and $\beta$, in addition to Stokes' number, $\mathcal{S}$, are also listed.}
\label{cnoidaltable}
\end{center}
\end{table}

This is a fourth-order polynomial in $H$, 
and by fixing a value for $m$ it can be solved numerically to obtain the 
maximum allowable wave height for the cnoidal wave, $H_{\text{max cnoidal}}(m)$. 
Since the periodic wave reduces to the solitary wave 
in the nonlinear limit ($m \rightarrow 1$), 
and $m \rightarrow 0^+$ is the linear limit, 
we search for real roots of \eqref{cnpol} in the interval $[0,1]$.

Different values of $m$ and corresponding values of $H_{\text{max cnoidal}}(m)$ are listed in Table \ref{cnoidaltable}.
The table also shows corresponding values of wave length, $\lambda$
and the dimensionless parameters $\alpha$ and $\beta$. In addition to Stokes' number $\mathcal{S}$, 
defined by the ratio $\alpha/\beta$ is tabulated.

\begin{figure}
\centering
\includegraphics[scale = 0.35]{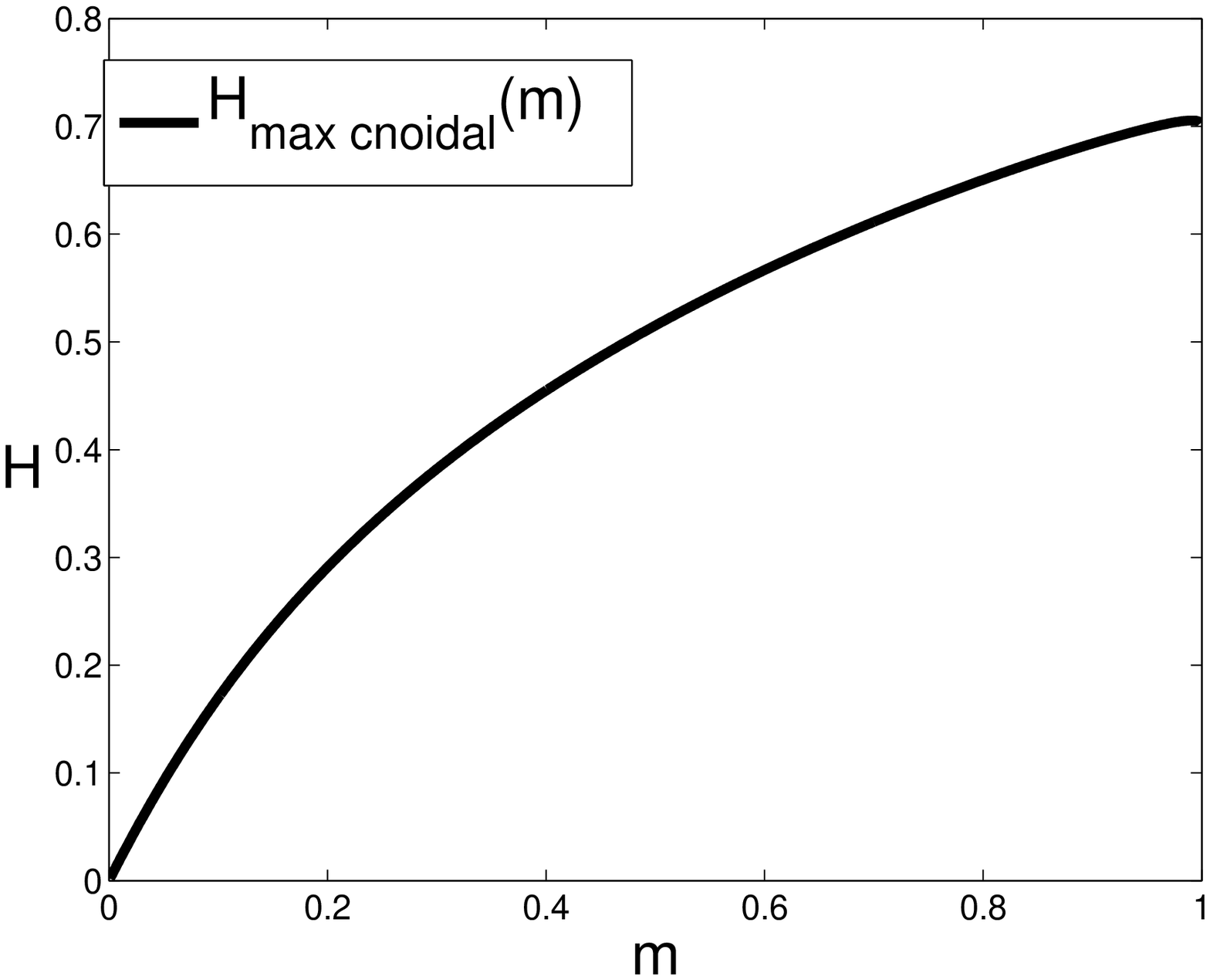}~~~~~~
\includegraphics[scale = 0.35]{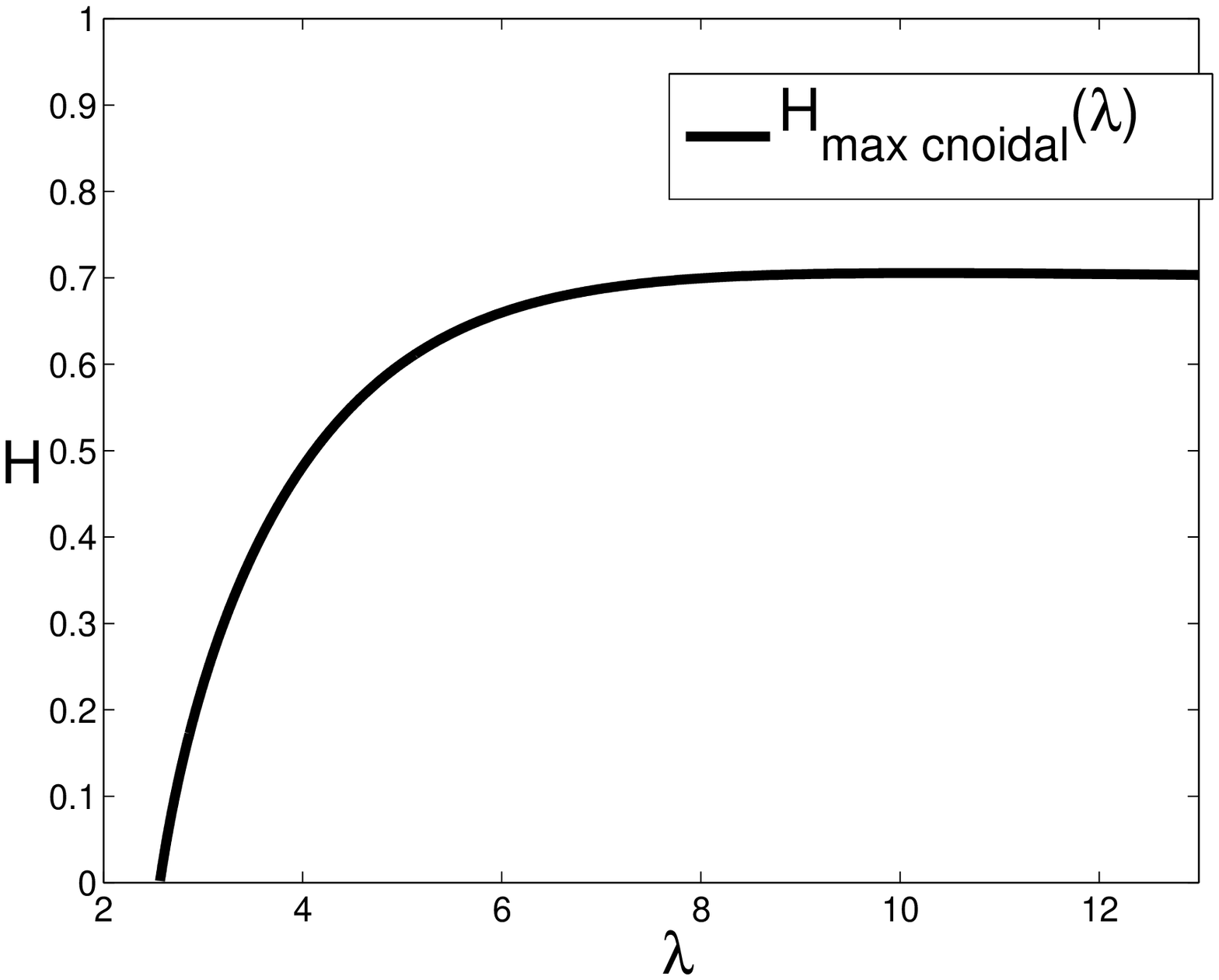}
\caption{\small Left: Maximum allowable wave height for the cnoidal solution as a function of $m$.
Right: Maximum allowable wave height for the cnoidal solution as a function of wavelength.}
\label{fig:Hcritvsm}
\end{figure}

Figure \ref{fig:Hcritvsm} shows a plot of $H_{\text{max cnoidal}}(m)$ 
obtained using a finer resolution in $m$ than displayed in the table above.
It can be seen that the maximum wave height seems to approach zero as $m$ approaches zero,
which fact is related to a going 'outside' of the domain of validity of the Boussinesq approximation, 
where $\alpha$ and $\beta$ are assumed to be of the same order of magnitude. 
It is also worth noting that the limiting values of $a$, $b$ and $c$ as $m \rightarrow 1$ are:
\begin{equation*}
a = 1, \ b = 0, \ \text{ and } \ c = 0
\end{equation*}
Applying this to equation \eqref{cnpol} gives the breaking criterion for the solitary wave, 
equation \eqref{solpol}. This is also easy to see if we plot the maximum wave height of 
the cnoidal wave as a function of wavelength $\lambda$.

\section{Numerical study of wave breaking in undular bores}
A bore or traveling hydraulic jump is the transition between two uniform flows with different depths.
Assuming that one of these flow depths is the undisturbed water level $h_0$,
and the incident water level is given by $h_0 + a_0$, we denote by 
$\alpha=a_0 / h_0$ the strength of the bore.
With the nondimensionalization explained in the introduction, the undisturbed depth is $1$,
and the incident water level is $\alpha$.
In this section we aim to find the threshold for which the bore transitions from being purely undular 
to one which exhibits breaking in the leading wave. This threshold is defined in terms of the 
ratio $\alpha$, and was found experimentally to be $\alpha = 0.28$ \cite{Favre}.

As already mentioned, if the kinematic breaking criterion is used in connection with 
phase-resolving numerical models, there is no difficulty in evaluating
the local wave and particle velocities.
When coupled with the Boussinesq ansatz for the horizontal velocity, the
convective criterion allowed for a somewhat accurate description of the breaking 
of the leading wave behind an undular bore \cite{BK2}. 
In this study, numerical integration of a Boussinesq system for undular bores
indicated  that breaking should occur if the bore strength is greater than $0.37$. 
While this result is qualitatively in the right direction, the critical bore strength
is overestimated by about $30\%$. In the following, it will be shown how this
result can be improved.

In order to find the critical bore strength in the numerical KdV model,
we will therefore use the value of $\alpha=0.25$ as a starting point, and then proceed with small 
increments until the breaking criterion is reached, i.e. when the horizontal particle speed 
of the fluid exceeds the phase speed. 
The initial position of the bore front will be set to be the origin, 
and the bore will then travel downstream in the positive $x$-direction. 
Initial data are given by
\begin{equation}
\eta_0 (x) = \Sfrac{1}{2} a_0 \big[ 1 - \tanh(kx) \big],
\end{equation}
where $k$ is a model parameter denoting the steepness of the initial bore slope.
The numerical approach to obtaining approximate solutions of \eqref{kdv} with
the appropriate boundary conditions is detailed in the appendix.

\begin{figure}
  \begin{center}
     {\includegraphics[scale=0.4]{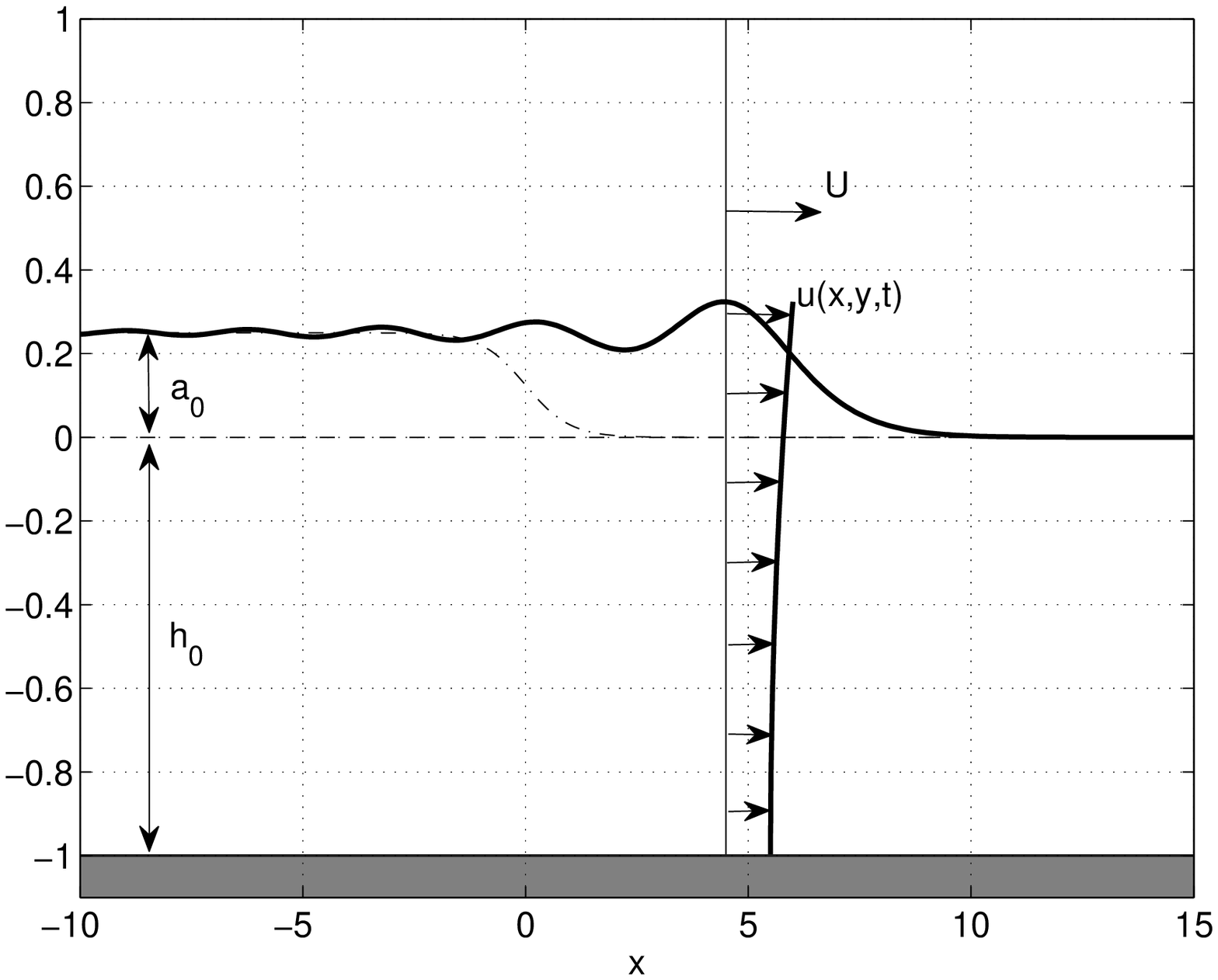}}
     {\includegraphics[scale=0.4]{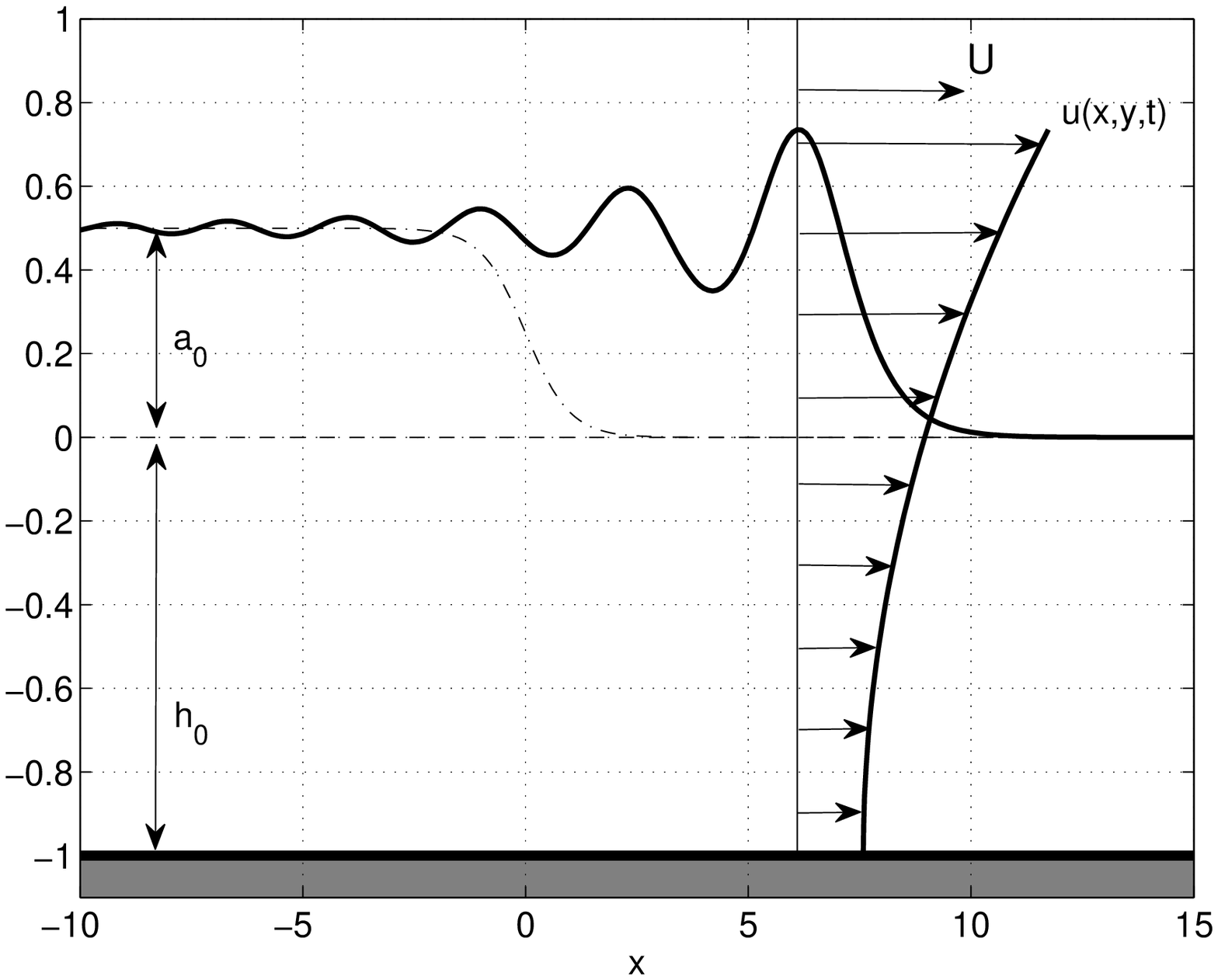}}
  \end{center}
  \caption{\small The surface profiles is given in the top panels and the
           horizontal velocities along the dashed lines below
           the wave-crests in the surface profiles is given in
           the bottom panels. The left panel is for nondimensional time $t=6$
           and the right for $t=6$. 
           The right-shifted initial surface profile
           data are indicated (dotted lines) in the top panel. 
           Left: Here $U=1.1$ is great than $u$ at the surface; no breaking.
           Right: Here $U=1.35$ is smaller than $u$ at the surface: breaking.}
\label{fig:boreprofiles}
\end{figure}

In order to test the convective breaking criterion \eqref{break1},
at each time step the horizontal particle velocity and the phase speed of the bore front
have to be calculated numerically. 
First, given the numerical solution $\eta(x_j,t_n) = \eta_j^n$,
we approximate the second spatial derivative of the surface, 
by the second-order central-difference formula
\begin{equation*}
\eta_{xx}(x_j,t) \approx \frac{\eta^n_{j-1} - 2\eta^n_j + \eta^n_{j+1}}{\delta x^2},
\end{equation*}
where $\delta x$ is the spatial grid size.
We then define the approximate horizontal particle velocity 
at a grid point $x_j$ and time $t_n$, and evaluated at the free surface 
as $u^n_j = u(x_j,\eta_j^n,t_n)$, such that equation \eqref{uab} yields
\begin{equation}
u^n_j = \eta_j^n - \Sfrac{1}{4} (\eta_j^n)^2 
+ \Big( \Sfrac{1}{3} + \Sfrac{(1+\eta_j^n)^2}{2} \Big) \Sfrac{\eta_{j-1}^n - 2\eta_j^n + \eta_{j+1}^n}{\delta x^2}.
\end{equation}

The phase speed of the first wave behind the bore front
can be approximated by measuring how the maximum of this wave propagates in time.
The left panel of Figure \ref{fig:boreprofiles} shows the initial profile 
with $a_0 = 0.25$ and $k = 1$ as a dashed curve, and the resulting bore profile
at $t=6$ as a solid curve. The figure also shows the horizontal component of the velocity
field underneath the first wavecrest, and it is apparent that this bore does not 
break as the particle velocity remains smaller than the propagation velocity of the leading wave.

After increasing $\alpha$ by increments of $0.001$ breaking is obtained for $\alpha_{crit} = 0.353$. 
A computation for $\alpha = 0.5$ (which is already past breaking) 
is shown in the right panel of figure \ref{fig:boreprofiles}.
It can be seen that the horizontal particle velocity is greater than the front
velocity after the solution has developed for some time.
As expected, the critical bore strength is bigger than the one found experimentally, 
but is fairly close to the one found in \cite{BK2}.


The maximum waveheight $H_{max}$ of the solution may be extracted 
at the particular time step where breaking first occurs.
If this waveheight
is compared to the critical waveheight of the solitary wave from section \ref{sec:breakingcrit},
it is found that the leading wave behind the bore breaks at a waveheight slightly
larger than the solitary wave. 
However, the values obtained for maximum wave height of the bore seem 
to approach the maximum height of the solitary wave as the bore strength decreases.
Since smaller bore strengths will have larger lead times before
the largest wave reaches the point of breaking,
in these cases, the leading wave has already separated as an individual
solitary wave, and the waveheight at breaking matches closely the 
value obtained for  $H_{\text{max solitary}}$.


The figure below shows a plot of the calculated values of $H_{\text{max}}$ as a function of time. 
A curve of the form $y_{\text{fit}} = At^B + C$ has also been fitted to the data.
\begin{figure}
\centering
\includegraphics[height = 4.8cm]{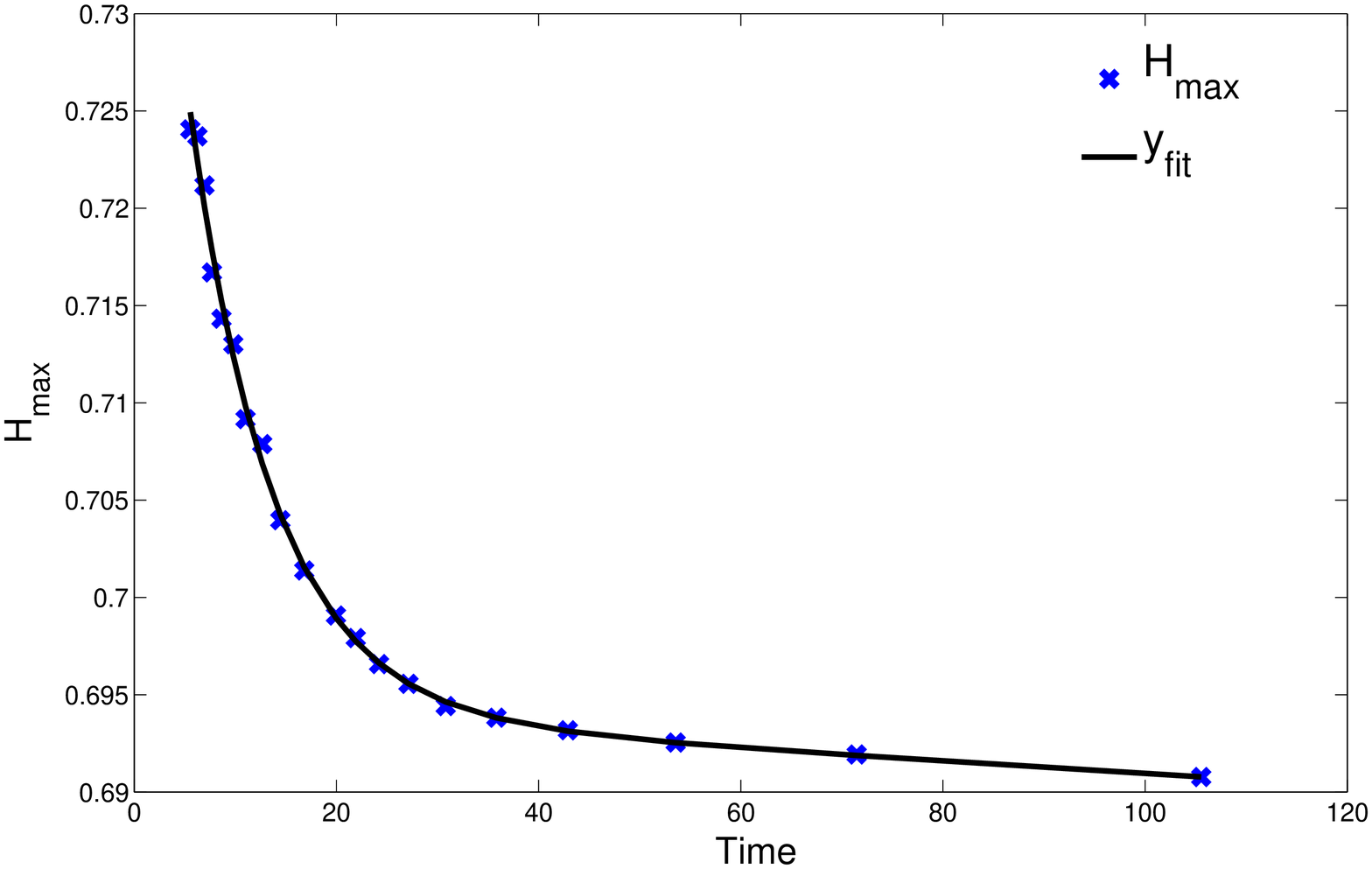}
\includegraphics[height = 4.8cm]{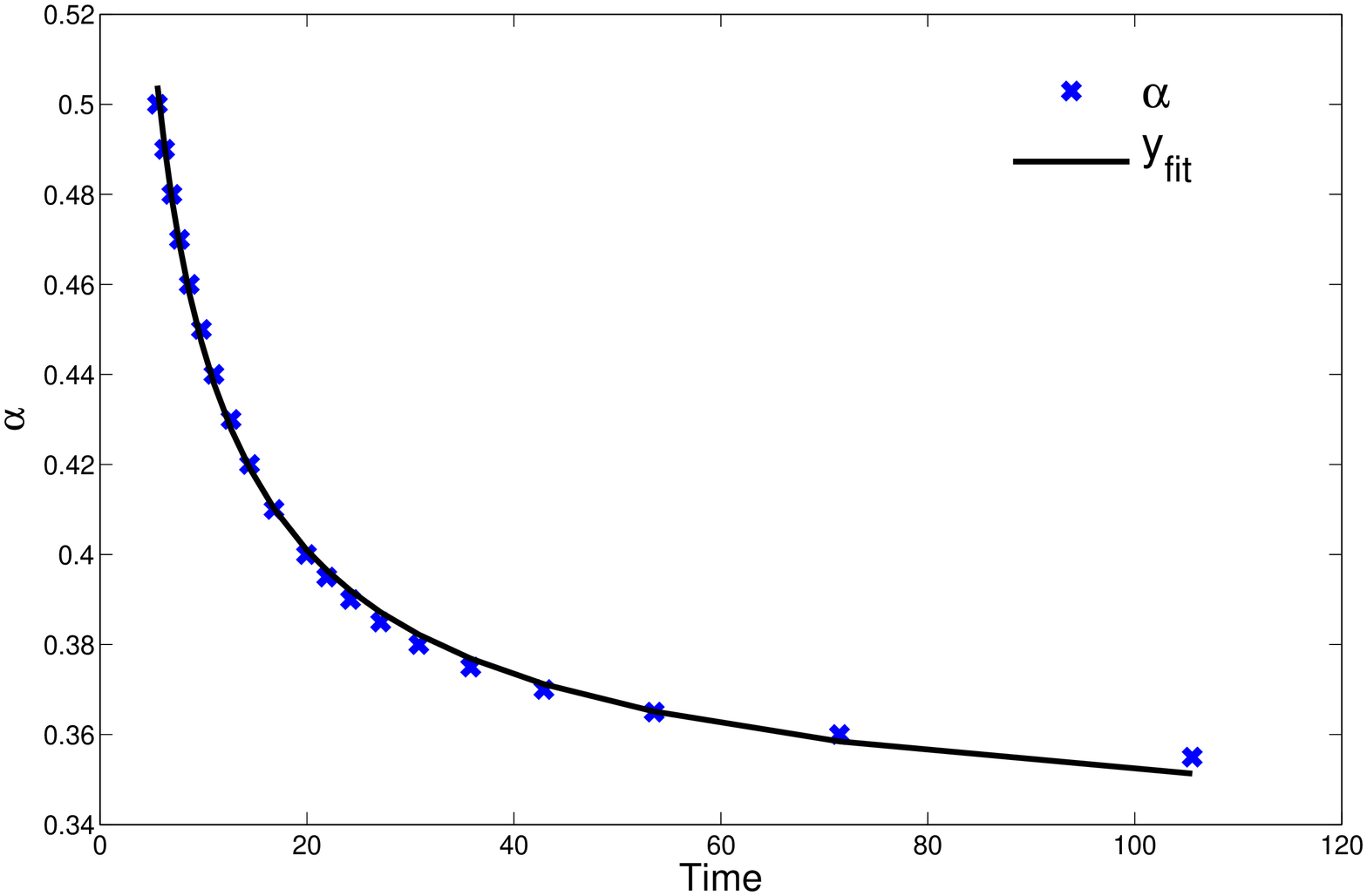}
\caption{\small Left panel: Maximum wave height as a function of time. Right panel:
Bore strength as a function of corresponding breaking times}
\label{BoreStrength}
\end{figure} 
The parameters $A$, $B$ and $C$ in the fitted curve are given by
\begin{equation*}
A = 0.1764, \ B = -0.8582 \ \text{ and } \ C = 0.6863,
\end{equation*}
and the residual sum of squares is $= 2.54\cdot 10^{-5}$.
The curve fits the data fairly well, and it is evident from the form of the equation 
that it approaches the value $c = 0.6863$ for large $t$.
This value is very close to the maximum possible height of the solitary wave
$H_{\text{max solitary}}=0.6879$.
We can therefore confirm that the bore eventually disintegrates into several solitary waves, 
but as the bore strength, $\alpha$, increases, the bore will break before this happens 
and thus cannot be described in terms of solitary waves. 
This observation also points to the possibility of obtaining a lower value of the critical
bore strength $\alpha_{\text{crit}}$ if a similar curve fit is used. 
To this end, we plot the bore strength as a function of corresponding breaking times.
If a curve of the same form as above is used, the parameters 
$A$, $B$ and $C$ in the fitted curve are given by
\begin{equation*}
A = 0.5751, \ B = -0.6908 \ \text{ and } \ C = 0.3283.
\end{equation*}
The residual is $= 7.74\cdot 10^{-5}$,
and visual inspection of Figure \ref{BoreStrength} indicates that the fitted curve fits 
the data rather well also in this case. The numerical value of $C$ 
indicates an asymptotic critical bore strength of $= 0.3283$ for large values of $t$, 
and this value is somewhat lower than the value of $\alpha_{\text{crit}}$ obtained previously.

\section{Conclusion}
It has been shown that the derivation of the KdV equation as a model for surface water waves
admits the definition of a local particle velocity field. The horizontal component of
this velocity field may be evaluated near the crest of a wave. For solitary and cnoidal
waves, evaluating the local Froude number leads to a convective wave breaking criterion
which shows that there is a limiting waveheight beyond which the waves feature 
incipient wave breaking. Thus even though the solitary wave and the cnoidal waves
are solutions of the KdV equation for any waveheight, they do not represent
valid approximate surface waves for waveheights beyond the critical waveheight.

The critical waveheight for solitary waves is $H_{\text{max solitary}} = 0.6879$
is quite large, and it is not surprising that these waves do not represent
reasonable approximations to real surface waves.
For the cnoidal waves, the situation is more complex. For waves near Stokes
number $1$, the limiting waveheight according to the convective breaking
criterion is in the range $0.15$ to $0.3$ (see Table 1), and these waves
are in the range where it is generally thought that solutions of the KdV
equation give a faithful approximation of real surface waves. 
As the elliptic parameter $m$ approaches $1$, the critical waveheight
of the cnoidal waves approaches  $H_{\text{max solitary}} = 0.6879$.
On the other
hand, the limit in the case when $m \rightarrow 0$ is the linear case
in which solutions have negligible amplitude, and the critical waveheight
in this case is indeed $0$. 

The convective breaking criterion has also been used to determine 
the critical bore strength $\alpha_{\text{crit}}$ 
for which an undular bore first exhibits breaking. 
The result $\alpha = 0.323$ is still somewhat higher than the experimentally 
determined value $\alpha = 0.28$, which was obtained using wave tank experiments \cite{Favre}, 
but lower than the ratio obtained using a similar numerical model in \cite{BK2}. 
The discrepancy with the experimental findings could be ascribed to
the assumptions made on the physical system, 
i.e. the absence of transverse motion of the fluid and an irrotational and inviscid flow.
While recent studies have highlighted the importance of these effects
\cite{Chanson,RG2013}, it is not clear that the inclusion of any of these
will improve the quantitative agreement between the current analysis
and the experimental findings of \cite{Favre}.
For example frictional damping due to molecular viscosity and boundary effects are more likely
to delay the onset of breaking than to facilitate it.
It therefore seems likely that the limiting procedure used in obtaining the model 
equations is responsible for some of the discrepancy, and a
possible avenue for improving the quantitative description of wave
breaking in an undular bore would be to include some of the higher-order terms
which were neglected in the KdV model. Such higher-order models
including both nonlinear and dispersive effects are well known
(see \cite{KRI2015} and the references therein).

Another possible improvement is the relaxation of the breaking criterion
recently advocated in \cite{BRB1,BRB2}.
In these works, several breaking criteria were used
in connection with numerical Boussinesq models,
and with the aim of quantifying the onset of breaking in shoaling waves.
It was found that
the convective criterion \eqref{break1} did not perform as well as other criteria based
on local waveheight \cite{Kazolea,Petti}, and a relaxation constant was introduced in order
to improve the analysis.
However, since this constant has to be determined empirically from a given set of experiments, 
we have chosen not to pursue such a development here.  \\

\noindent
{\bf Acknowledgments:}
This work was supported in part by the Research Council of Norway 
through grant no. 213474/F20.

\appendix

\section{The numerical scheme for the KdV equation}

The numerical approximation of solutions of the equation \eqref{kdv} is
based on a finite-difference method for the spatial derivatives
with appropriate boundary conditions,
and a hybrid Adams-Bashforth / Crank-Nicolson time integration scheme.
Since the KdV equation is of third order in spatial derivatives, 
three boundary conditions are needed for the numerical scheme. 
In the far field downstream of the bore front the surface elevation approaches zero, 
while in far field upstream it approaches $\alpha$. 
These far field conditions are chosen to be the boundary conditions which are exact 
up to machine precision as long as the spatial domain is large enough. 
Thus care has to be taken to ensure this is the case. 
For the third boundary condition a homogeneous Neumann condition is chosen on the right. 
To summarize, the problem to be solved is given by
\begin{equation}\label{nonhom}
\begin{split}
\eta_t + \eta_x + \Sfrac{3}{4} \big( \eta^2 \big)_x + \Sfrac{1}{6} \eta_{xxx} =& \ 0,\ x \in [-l,l], \ t \geq 0\\
\eta(x,0) = & \ \eta_0 (x),\\
\eta(-l,t) = & \ \alpha,\\
\eta(l,t) = & \ 0,\\
\eta_x(l,t) = & \ 0.
\end{split}
\end{equation}
In order to incorporate non-homogeneous boundary conditions, we define the auxiliary function
\begin{equation}\label{xi}
\xi(x,t) \equiv \eta(x,t) - \eta_0(x).
\end{equation}
Upon substituting for $\eta(x,t)$ in \eqref{nonhom} we have the problem in terms 
of $\xi$ and $\eta_0$ which now has homogeneous Dirichlet conditions along with 
the same homogeneous Neumann condition on the right. The problem \eqref{nonhom} now reads
\begin{equation*}
\xi_t + \xi_x + \Sfrac{3}{4} (\xi^2)_x + \Sfrac{3}{2}(\eta_0 \xi)_x + \Sfrac{1}{6} \xi_{xxx} =  \ -F, \ x \in [-l,l], \ t \geq 0,
\end{equation*}
where $F \equiv \eta_0' + \frac{3}{2} \eta_0 \eta_0' + \frac{1}{6} \eta_0'''$,
and homogeneous boundary and initial conditions are imposed.

To approximate the spatial derivatives a finite difference scheme is applied, 
while to advance in time, an explicit Adams-Bashforth method is used on the nonlinear terms 
and an implicit Crank-Nicolson method is used on the linear terms. 
Both methods are of second order, and while this does not automatically yield a 
second-order scheme for the full equation, numerical experiments in \cite{Skogestad2009} 
suggest that this is in fact the case.

We continue by discretizing the spatial domain, $[-l,l]$ 
uniformly using a finite set of points, $\{x_j\}_{j = 0}^{N} \subset [-l,l]$, 
where $x_0 = -l$ and $x_N = l$, and $\delta x = 2l/N$ 
is the distance between two neighboring grid points. 
The time domain is also discretized uniformly using $t_n = n\delta t$, where $t_0 = 0$. 
We define the approximate function value at time $t_n$ and grid point $x_j$ 
to be $v_j^n = \xi(x_j,t_n)$. 
The solutions $\eta(x_j,t_n) = \eta_j^n$ is then given by \eqref{xi}.

The first and third derivatives at a point $x_j$ are approximated as follows using the central difference formulas:
\begin{equation*}
\begin{split}
\xi_x(x_j,t) &\approx \frac{v_{j+1} - v_{j-1}}{2\delta x}\\ 
\xi_{xxx}(x_j,t) &\approx \frac{v_{j+2} - 2v_{j+1} + 2v_{j-1} - v_{j-2}}{2\delta x^3}\\
\end{split}
\end{equation*}
From the Dirichlet conditions we have that $v_0 = 0$ and $v_N = 0$ 
so we need only solve the equation for the grid points $\{x_j\}_{j = 1}^{N - 1}$. 
This leaves us with only two points for which the third derivative approximation is not valid.
From the Neumann condition we have that $\xi_x(l,t) = 0$, 
so by writing this as a central difference approximation we get that
\begin{equation*}
\frac{v_{N+1} - v_{N-1}}{2\delta x} = 0 \implies v_{N+1} = v_{N-1}.
\end{equation*}
This enables us to use the third derivative approximation at grid point $x_{N-1}$ as follows
\begin{equation*}
\xi_{xxx}(x_{N-1},t) \approx  \frac{ v_{N+1} - 2v_{N} + 2v_{N-2} - v_{N-3}}{2 \delta x^3} = \frac{v_{N-1} + 2v_{N-2} - v_{N-1}}{2\delta x^3}.
\end{equation*}
As there is no Neumann condition on the left we use a forward difference formula to approximate 
the third derivative at grid point $x_1$:
\begin{equation*}
\xi_{xxx}(x_1,t) \approx \frac{-v_4 + 6v_3 - 12v_2 + 10v_1 - 3v_0}{2\delta x}.
\end{equation*}
Defining difference matrices in this way, we may set up the following difference equation:
\begin{equation*}
\frac{\mathbf{v}^{n+1} - \mathbf{v}^n}{\delta t} = - \frac{3}{4} D_1 \big( \mathbf{v}^n \big)^2 - \frac{3}{2} D_1 \mathbf{v}^n \mathbf{\eta_0} - \frac{1}{6} D_3 \mathbf{v}^n  -D_1 \mathbf{v}^n - \mathbf{F},
\end{equation*}
where $\mathbf{v}^n = (v_1^n, v_2^n, ..., v_{N-1}^n),\ \mathbf{\eta_0} = (\eta_0(x_1), \eta_0(x_2), ...,\eta_0(x_{N-1})) \\$ 
and $\mathbf{F} = (F(x_1), F(x_2), ...,F(x_{N-1}))$.

By applying the Adams-Bashforth method to the two first terms on the right, 
and the Crank-Nicolson method to the next two terms we get
\begin{equation*}
\begin{split}
\frac{\mathbf{v}^{n+1} - \mathbf{v}^n}{\delta t} = -& \ \frac{3}{4} D_1 \bigg[ \ 3 \Big( \Sfrac{1}{2} \big( \mathbf{v}^n \big)^2 +  \mathbf{v}^n \mathbf{\eta_0} \Big) - \Big( \Sfrac{1}{2} \big(\mathbf{v}^{n-1} \big)^2 
    +  \mathbf{v}^{n-1} \mathbf{\eta_0} \Big) \ \bigg] \\
-& \ \frac{1}{2} \bigg( D_1 \Big[ \mathbf{v}^{n+1} + \mathbf{v}^n \Big] 
+ \Sfrac{1}{6} D_3 \Big[ \mathbf{v}^{n+1} + \mathbf{v}^n \Big] \ \bigg) - \mathbf{F}.
\end{split}
\end{equation*}
At each time step this equation has to be solved for $\mathbf{v}^{n+1}$ to advance in time. 
By defining the matrix $E = (I + \frac{\delta t}{2} D_1 + \frac{\delta t}{12} D_3)$ for convenience, 
we do this as follows ($I$ denotes the $(N-1)\times(N-1)$ identity matrix):
\begin{equation*}
\begin{split}
\mathbf{v}^{n+1} =& \ E^{-1} \Big[ I - \Sfrac{\delta t}{2} D_1 - \Sfrac{\delta t}{12} D_3 \Big] \mathbf{v}^n \\
&-  \frac{3 \delta t}{4} E^{-1} D_1 \bigg[ \ 3 \Big( \Sfrac{1}{2} \big(\mathbf{v}^n \big)^2 
      +  \mathbf{v}^n \mathbf{\eta_0} \Big) - \Big( \Sfrac{1}{2} \big(\mathbf{v}^{n-1} \big)^2 
    +  \mathbf{v}^{n-1} \mathbf{\eta_0} \Big) \ \bigg]  - \delta t E^{-1} \mathbf{F}.
\end{split}
\end{equation*}
This method requires function values at two previous time steps to calculate the next one, so at the very first iteration a different approach is needed. One way of fixing the problem is to use a forward Euler method on the nonlinear terms at the first time step. This method has a local truncation error of order two and any instability issues will not be a problem for one single time step. Thus we get the following difference equation for the first time step:
\begin{equation*}
\frac{\mathbf{v}^{2} - \mathbf{v}^1}{\delta t} 
= -  \frac{3}{2} D_1 \Big[ \Sfrac{1}{2} \big( \mathbf{v}^1 \big)^2 + \mathbf{v}^1 \mathbf{\eta_0} \Big]
- \frac{1}{2} \bigg(  D_1 \Big[ \mathbf{v}^{2} + \mathbf{v}^1 \Big] 
               + \Sfrac{1}{6} D_3 \Big[\mathbf{v}^{2} + \mathbf{v}^1 \Big] \ \bigg) 
- \mathbf{F}.
\end{equation*}

\end{document}